# On-the-Fly Path Planning for the Design of Compositional Gradients in High Dimensions

Samuel Price[1], Zhaoxi Cao[1], Ian McCue[1*]

[1]*Department of Materials Science and Engineering, Northwestern University, Evanston, IL 60208, USA*

*\*Corresponding author: ian.mccue@northwestern.edu*

**Abstract**

Functional gradients have recently experienced an explosion in activity due to advances in manufacturing, where compositions can now be spatially varied on-the-fly during fabrication. In addition, modern computational thermodynamics has reached sufficient maturity – with respect to property databases and the availability of commercial software – that gradients can be designed with specific sets of properties. Despite these successes, there are practical limitations on the calculation speeds of these thermodynamic tools that make it intractable to model every element in an alloy. As a result, most path planning is carried out via surrogate models on simplified systems (e.g., approximating Inconel 718 as $Ni_{59}Cr_{23}Fe_{18}$ instead of $Ni_{53}Cr_{23}Fe_{18}Nb_3Mo_2Ti_1$). In this work, we demonstrate that this limitation can be overcome using a combination of on-the-fly sampling and a conjectured corollary of the lever rule for transformations of isothermal paths in arbitrary compositional dimensions. We quantitatively benchmark the effectiveness of this new method and find that it can be as much as $10^6$ times more efficient than surrogate modeling.

# 1 Introduction

Operating conditions are becoming increasingly more complex across numerous engineering applications, but it is most apparent in the aerospace, energy, and transportation sectors. Property requirements for components in these applications vary substantially along their dimensions, often necessitating completely different material classes from one end to another. These dissimilar materials are frequently connected via mechanical joints and fasteners, which add unnecessary weight and have sharp property transitions. However, mechanical joints are currently preferred because chemical interactions between dissimilar materials can lead to new phases that compromise performance. Ideally, the transition between two distinct materials would be continuous without any deleterious phases present.

Some successful joining recipes have been identified through experimental guess-and-check, but the development of more intentional approaches has been ongoing since the 1940s [1]. Moving to the present, composition gradients are commonly employed to transition between dissimilar materials [2–8]. Equilibrium phase diagrams often aid design by highlighting regions where undesirable phases can form and how they connect to certain base alloy compositions [9]. This approach has been recently extended, through the coupling of CALculations of PHAse Diagram (CALPHAD) software and path planning algorithms, to construct composition gradients in arbitrary element systems [10]. These gradients can be specified to not only satisfy equilibrium phase constraints but also to optimize a particular cost function (e.g., path length, clearance from bad phases, thermal expansion mismatch, etc.) [10–12].

The most versatile of these path planning algorithms are the sample-based planners, namely the rapidly-exploring random tree (RRT) algorithm and its variants [13,14]. These algorithms stochastically sample the configuration space and attempt to connect points that satisfy the imposed constraints, building out a tree of valid connections. Sample-based planners rely on information from a collision-checking function, which maps inputs in the configuration space to a simple classification scheme (feasible or infeasible) depending on whether they satisfy the imposed constraints. The feasibility of a composition is typically defined by imposing constraints on the phases present at that composition, which is often done by classifying certain phases (and their fractions) as "good" or "bad"; these classifications are based on whether the phases have the potential for favorable or unfavorable physical properties. In this work, the phases associated with a composition will be the CALPHAD predicted equilibrium phases for that composition at a fixed temperature (and atmospheric pressure). However, it should be noted that others have explored synthesizing phase information across a range of temperatures either through direct equilibrium calculations [10] or via Scheil-Gulliver calculations [15,16].

In designing composition gradients with sample-based planners, prior works in the literature have used machine-learning to construct a surrogate model of the "ground truth" collision function [10,11,17].

These models summarize the CALPHAD information they are trained on and can be used to estimate relevant properties of input compositions at speeds orders of magnitude faster than evaluating the ground truth function (CALPHAD). However, the use of a surrogate model only reduces the overall path planning time for low-dimensional systems (≤ 4 elements) because the number of ground truth samples needed to construct a surrogate model grows rapidly with the number of elements in the system. The exact scaling depends on the specific model and sampling scheme used, but it is typically exponential in the number of elements (dimensions) with one example using at least $5\times10^d$ samples for a system of $d$ elements [17]. By contrast, the number of collision checks an RRT needs to find a feasible path does not explicitly depend on the dimension but rather on the "visibility" of the free space [18]. In fact, it will be shown later in this manuscript that the fraction of the total system that must be sampled to find the first feasible path decreases with the dimension of the system and building a surrogate model becomes an increasingly inefficient strategy.

Given this inversion at higher dimensions, the manuscript at hand foregoes surrogate modeling of bad phase regions when planning feasible paths in many-element systems (> 4 elements). Instead, we opt to perform collision checks by sampling the ground truth function (i.e., CALPHAD) directly. We find that this approach leads to a reduction in the time to find a viable path between two alloys, using four structural alloys common to AM. In addition, the number of CALPHAD queries can be further reduced – by orders of magnitude – by increasing the visibility of the free space. This increased visibility comes at the cost of allowing high fractions of deleterious phases in the gradient. However, we address this limitation via a phase-boundary conjecture, which enables any "partially bad" isothermal path to be transformed out of deleterious regions using a formula that only requires information from the existing CALPHAD calculations. Taken together, this work represents a substantial development in the utility of RRT algorithms for path planning in high dimensions. A schematic overview of our method and how it compares with previous approaches in the literature is shown in Figure 1.

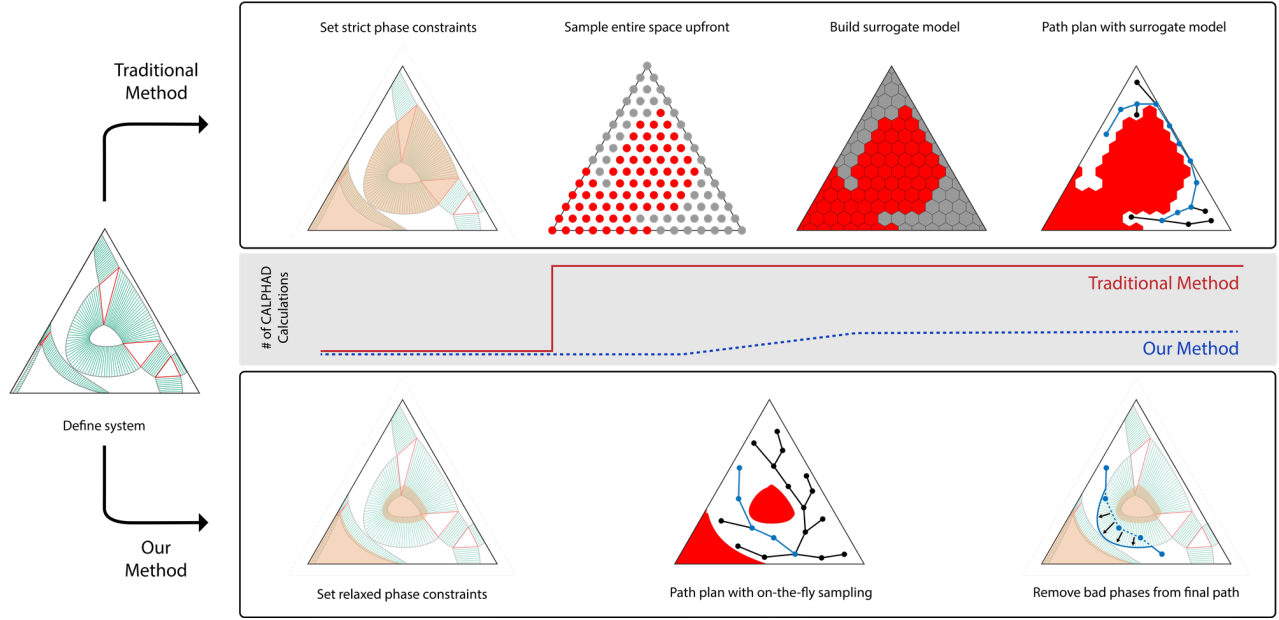

*Figure 1: A schematic outlining the traditional surrogate-model based approach (upper row, taking inspiration from Figure 2 of [11]) and our on-the-fly method for composition gradient design (lower row). The traditional workflow involves sampling the entire composition space upfront to construct a surrogate model to assist in path planning. This upfront sampling requires a large number of CALPHAD calculations and therefore time (red line). By contrast, our method samples only the number of CALPHAD calculations demanded by the path planning algorithm in real time (blue line), which is often significantly less than needed for the surrogate model construction[\*]. Additionally, by first planning around relaxed phase constraints (e.g., allowing up to 80% total bad phase) our approach greatly improves the efficiency of path planning. The final path can then be transformed out of these partially bad regions by leveraging a newly conjectured property of phase diagrams. The use of relaxed phase constraints and direct CALPHAD sampling enables our method to find new, and perhaps more desirable, paths than the traditional approach.*

## 2   Methods

The CALPHAD calculations used in this framework are single equilibrium calculations performed by Thermo-Calc using the TC-Python module and TCFE9 database [19]; a constant database was chosen for the sake of evaluating the methods in this paper, but it would be more accurate to use a specific database tailored for each alloy systems of interest. We use a rapidly-exploring random tree (RRT) as our path planning algorithm. Although several variations have been developed, including RRT* [13] and RRT*FN [20], unless stated otherwise we use the original formulation of the algorithm [14]. The selection of an appropriate steering distance is important to minimize the number of CALPHAD queries as selecting either too large or too small of a steering distance can both lead to significant inefficiencies. For the alloy space explored here, we used $\frac{\sqrt{2}}{10}$, representing 10% of the maximal distance within any composition space (regular simplex). Our resolution for discretizing edges during collision checks was 0.01. The edge collision

---

[\*] Although the figure illustrates these approaches on ternary phase diagrams (due to the limitations of a 2D figure) our approach is most beneficial for systems of 4+ elements (see Section 3.4 Dimensional Scaling).

checking strategy first tested the midpoint and then check the remaining points only if the midpoint was collision free. Calculations were carried out on a Dell workstation with 64GB of RAM and an Intel Xeon Silver 4214R, 2.40GHz processor.

The system is defined once at the beginning of a given RRT run using the entire combined element set. For each subsequent composition tested, a new "SingleEquilibriumCalculation" is defined from the system to prevent prior results influencing future ones. For each calculation, the thermodynamic results are extracted using a modified version of the "list_stable_phases" function. The equilibrium phase names and their respective mole fractions are then used to determine whether the composition in question satisfies our phase constraints. The single equilibrium calculations use global minimization and start with the default value of 2000 max_grid_points. If a calculation fails to converge it is rerun with max_grid_points=20,000. If this too fails one final attempt is made with max_grid_points=200,000. If this fails, we give up in order to save time and prevent excessive memory consumption. To be conservative, we assume the point fails the constraints.

## 3   Results & Discussion

This work focuses on the problem of finding the first feasible isothermal path (i.e., one that avoids bad phase regions) between two start/end points in the composition space formed by the union of the elements in each alloy. Our test case consists of composition pairs from a set of four commonly used alloys (IN718, Ti64, 316L, AlSi10Mg) with path planning done at 400°C. To this end, we use the simplest classification scheme and declare FCC, BCC, and HCP phases as good, non-good phases that are present in the base alloys are considered "limited" phases, and all others (liquid, LAVES, etc.) as bad. The time needed to find the first feasible path in the composition space between two alloy compositions represents a minimum in terms of useful path planning. It will be shown at the end of this manuscript that this approach is also able to carry out optimal path planning.

### 3.1   On-the-fly Sampling Methods for High Dimensions

As noted above, the use of a surrogate model for collision checking necessitates a minimum number of upfront ground truth evaluations (GTEs) via CALPHAD. As this number scales strongly with the dimension of the system, path planning with a surrogate model in a many-element system (>4 elements) becomes untenable (>>1 week). In contrast, the number of collision checks that must be made by an RRT to find the first feasible path in a many-element system is, as we have found, often significantly less than the number needed for surrogate model construction. For this reason, we opt to directly query the ground truth (CALPHAD) function itself, when performing collision checks, to find a viable path much more rapidly than the time needed for surrogate model construction. We refer to this as the on-the-fly (OTF) sampling method.

An illustrative example comparing the OTF method to surrogate modeling is shown in Figure 2, where a path is constructed between Ti64 and AlSi10Mg in five dimensions – the lowest number of elements considered in this study. Figure 2(a&b) showcases an optimal path generated with RRT* [13] and a surrogate model (k-nearest neighbors classifier with k=1), where the cost function is path length (in composition space) which results in a minimal length path. However, this surrogate model required ~4.6 million CALPHAD calculations to be constructed. In contrast, the path in Figure 2(c&d) was generated via the OTF method, which only needed 3,222 CALPHAD queries (collision checks).

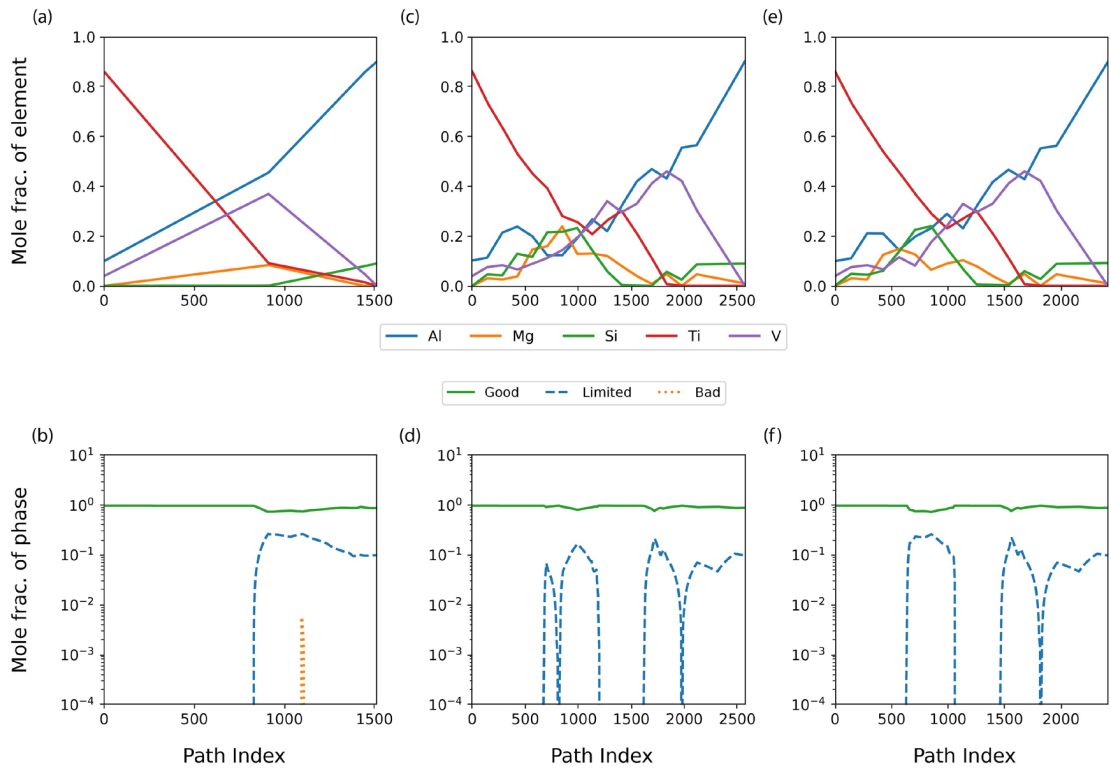

*Figure 2: Comparison of paths from Ti64 to AlSi10Mg using three different computational methods. (a), (c), (e) are the elemental composition profiles; and (b), (d), (f) are the predicted equilibrium phase fractions summed by classification, where the total amount of "good" phases is shown in green, "limited" phases (i.e., secondary phases in these alloys used for strengthening) in blue, and "bad" phases in orange. (a), (b) is a path optimized to minimize path length using surrogate model queries, whereas (c), (d) is the first feasible path found using OTF sampling, and (e), (f) is the first feasible path using surrogate model queries. Each path avoids deleterious phase regions, but the optimized path has a shorter path length in composition space, 1.51, than either first feasible path, 2.41 and 2.57 for (c) and (e) respectively.*

The composition profile in Figure 2(c&d) is the first feasible path, which globally takes a similar path to (a&b) but has more slope changes locally due to the nature of how the RRT algorithm selects points and makes connections. Thus, the abrupt composition changes along the OTF path are not as "smooth" as the one in Figure 2(a&b), which is able to be refined through subsequent inexpensive collision checks (albeit at the penalty of 3 orders of magnitude more CALPHAD queries to generate the surrogate model). A more appropriate comparison to the OTF method is shown in Figure 2(e&f), where the surrogate model was queried 3,225 times and produced a similar path to that made via the OTF method. For some compositions, the surrogate model classification differs from the true classification, which explains why the paths in Figure 2(c&e) are not identical and the number of queries made are slightly different.

Using our OTF method, the runtime remains dominated by the number of GTEs required, but *the GTEs are now the result of collision checks made by the RRT rather than upfront sampling for surrogate model construction*. Each collision check is an expensive GTE – in contrast to the cheaper surrogate model evaluation – but this outcome is mostly advantageous. The number of collision checks needed to determine the first feasible path (in high dimensions) is far fewer than the number of GTEs for surrogate model construction. While the OTF approach solves the issue of untenable surrogate model construction time, for many systems of interest the runtime is still too long (i.e., several days). However, this change in rate-limiting GTEs presents an opportunity to further reduce the time taken to find the first feasible path by reducing the number of necessary collision checks.

A major benefit of using an RRT algorithm is that the number of collision checks needed to find the first feasible path does not explicitly depend on dimension, but rather on the "visibility" properties of the free space [18]. Generally, the narrower the corridors in the free space that must be navigated the more samples the RRT will need to find such a feasible path. In many-element systems, it is unlikely that all elements will have a wide solubility range in the "good phase" solid solutions and will instead "crash out", forming deleterious phases. This point is especially true when potent intermetallic forming elements like Al, Si, and Mg are present. Limited solubility renders many of the good phase regions "narrow" (in at least one dimension), which results in small hypervolumes that require extensive sampling to be navigated by an RRT.

Increasing the total number of elements in the system increases the likelihood of such solubility range issues, and severely limits the utility of employing an RRT algorithm in high dimensions. Subspace sampling methods have proven effective at handling these solubility issues if they occur where the concentrations of the problematic elements are near zero [17]. This method is incorporated into our OTF framework. However, there still exist many instances in the *interior* of the composition space where the visibility of the free space is greatly restricted by the lack of solubility range. Thus, subspace inclusive sampling alone is not sufficient to overcome this problem.

## 3.2 Improving Visibility of RRT Algorithms

One method to increase the visibility of the free space is to relax the feasibility criteria by increasing the "allowable amount of total fraction of bad phase" from 0 to a value <1. Following the Lever Rule, the volume of the free space will grow monotonically with the total fraction of bad phase allowed. By allowing higher fractions of bad phases, narrow corridors are now widened and the number of collision checks needed for the RRT to find a feasible path greatly decreases. This effect is most pronounced for systems with initially very poor visibility (such as 316L SS or IN718 to AlSi10Mg).

We refer to this as the "relaxed constraint" RRT, and the feasible path it produces is likely to have a significant fraction of bad phases along it. However, provided the path does not pass through regions where the total bad phase fraction is unity, we posit that this "partially-bad" path can be transformed into a path with zero bad phase fraction at every point. Furthermore, this transformation, which we will refer to as the Bad Phase Purge (BPP) method, can be performed with minimal to no additional CALPHAD calculations. While the details of the BPP conjecture are provided in the Appendix, we will summarize a few key points here. The mathematical transformation is given by:

$$\sigma^*(\alpha) = H_2\big(\sigma(\alpha)\big) \left[ \frac{1}{\big(H_1(\sigma(\alpha)) \cdot \vec{V}_{good}\big)} \big(H_1(\sigma(\alpha)) \odot \vec{V}_{good}\big) \right] \quad (1)$$

where: $\sigma$ is the composition along the original path; $\sigma^*$ is the composition along the transformed path; $\alpha$ parameterizes the position along the path, where 0 is the start and 1 is the end; $\vec{V}_{good}$ identifies if a phase is good (1) or bad (0); $H_1$ maps the overall composition to phase fractions; and $H_2$ maps the overall composition to a matrix containing the composition of each phase.

Figure 3 illustrates the transformation of a path using the BPP method by using a ternary slice of the Co-Fe-Mn system at 900 K as a test case. This system has little significance beyond having two deleterious phases (SIGMA and A13) that occupy large fractions of the ternary diagram. These deleterious phases have finite composition ranges, but their limited solubility in the nearby favorable phases (FCC and BCC) lead to large phase coexistence regions. If we impose a strict constraint on phases within the gradient (i.e., zero fraction of deleterious phases), the obstacles for path planning are now denoted by the orange shapes in Figure 3.

Because these orange obstacles occupy most of the Gibbs triangle, the RRT algorithm is forced to plan through one of two narrow corridors. Normally, the RRT would require a large number of samples to successfully navigate this space. However, if the constraint is relaxed (here, any point along the path can contain <80% of a deleterious phase) a path between these two points is found using very few CALPHAD calculations. This path $\sigma$ is obviously not desirable due to the high SIGMA phase fractions but can be transformed to $\sigma^*$ using Equation (1). The initial and transformed paths (gray and cyan, respectively) are discretized in Figure 3 since we can only evaluate $H_1$ and $H_2$ by performing a CALPHAD calculation.

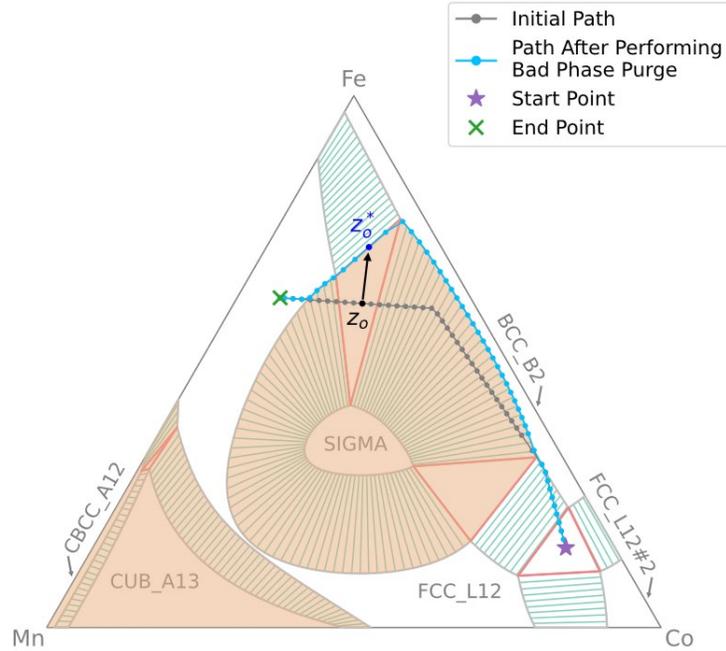

*Figure 3: Illustration of the Bad Phase Purge operation being applied to an arbitrary path in the Co-Fe-Mn system at 900 K. The initial path, $\sigma(\alpha)$, is shown in gray. A significant portion of it passes through phase regions containing the undesirable SIGMA phase, highlighted in orange. This path is transformed in a pointwise manner by the Bad Phase Purge operation to yield a new path free of any bad phases, $\sigma^*(\alpha)$, shown in cyan. One point, $z_o$, is highlighted along with its transformed result, $z_o^*$, to demonstrate the details of the transformation.*

The mathematical formulation in Equation (1) leverages the Lever Rule in an arbitrary dimension. Thermodynamically, multiphase regions possess the constraint that the chemical potential of individual species (elements) is equal in every phase present. Thus, any composition lying in a multiphase region decomposes into the respective equilibrium phases each with a specific composition. These phase compositions form a tie-simplex (generalization of a tie line) [21]. Moving within this tie-simplex does not change the composition of the equilibrium phases, but rather their relative fractions. Taking advantage of this thermodynamic property, we can "purge" any deleterious phase from the path by moving along the tie-simplex towards the good phase compositions and away from the bad phase compositions. In this example, ordered FCC and BCC phases were considered favorable, but this definition could change depending on the problem at hand. These phases could have been identified as problematic; however, doing so would ruin the utility of this example because no path would exist in this low-dimensional space.

While the entire path is transformed in Figure 3, we will show the transformation of a single point

$$z_o = \sigma(\alpha_o) = \begin{bmatrix} 0.21 \\ 0.61 \\ 0.18 \end{bmatrix} \begin{matrix} - \text{Co} \\ - \text{Fe} \\ - \text{Mn} \end{matrix}. \quad (2)$$

To complete Equation (1), we have the following additional inputs:

$$\vec{V}_{good} = \begin{bmatrix} 1 \\ 0 \\ 0 \\ 1 \\ 1 \\ 0 \end{bmatrix} \begin{matrix} - \text{BCC\_B2} \\ - \text{CBBC\_A12} \\ - \text{CUB\_A13} \\ - \text{FCC\_L12} \\ - \text{FCC\_L12\#2} \\ - \text{SIGMA} \end{matrix}, \quad H_1(z_o) = \begin{bmatrix} 0.34 \\ 0 \\ 0 \\ 0.30 \\ 0 \\ 0.36 \end{bmatrix}, \quad H_2(z_o) = \begin{bmatrix} 0.19 & 0 & 0 & 0.14 & 0 & 0.28 \\ 0.77 & 0 & 0 & 0.66 & 0 & 0.42 \\ 0.04 & 0 & 0 & 0.20 & 0 & 0.3 \end{bmatrix} \quad (3)$$

Substituting Equations (2) and (3) in (1), we find the transformed point:

$$z_o^* = \begin{bmatrix} 0.17 \\ 0.72 \\ 0.11 \end{bmatrix} \quad (4)$$

, which can be readily identified in Figure 3 as a composition that exists in the L12/B2 two-phase region.

## 3.3 Evaluation on Test Case with Four Commonly Used Alloys

We can now evaluate the effectiveness of using the BPP approach by finding feasible paths between all pairs of four commonly used structural alloys: 316L SS, Ti64, IN718, and AlSi10Mg. The paths are planned at an arbitrary isothermal temperature of 400°C using the TCFE9 database as it contains all the necessary elements [19]. FCC, HCP, and BCC phases are considered "good"; non-good phases that are present in the base alloys at 400°C are considered "limited" phases (they must be considered non-bad so that the start/end points of the path are valid, but their amount can be limited); and all other phases are considered "bad". These alloys are summarized in Table 1.

To properly quantify the OTF methods developed in this paper, we applied both a strict RRT and a relaxed RRT (with BPP post-processing) to all six pairs. It should be noted that the existence of such feasible paths is not guaranteed and was not known to the authors prior to conducting these tests. For the relaxed constraint, paths consisting of entirely "good" points were found in every instance, after applying the BPP. For the strict constraint, successful paths were only found in four pairs; the other two pairs (316L-AlSiMg and IN718-AlSiMg) were run as long as possible but stopped due to the unlikelihood of finding a path within a reasonable time scale (less than two weeks).

Figure 4 shows the equilibrium phase fractions as a function of the normalized distance along the path for one random seed; the corresponding composition profiles can be found in Supplementary Figure 1. The amount of total bad phase is effectively zero ($<10^{-2}$) at every point, while the limited phases only appear at the ends of the gradients$^\dagger$ and in fractions less than or equal to their amount in the base alloys. Before the RRT starts in earnest, the linear composition profile between the two compositions is tested. If it is feasible, then the run stops and returns the linear path, otherwise the RRT proceeds as normal. At 400°C, only the 316L SS to IN718 system has a valid straight-line path, which explains the simplicity of its phase fraction plot.

*Table 1: Description of the base alloys used for framework evaluation*

| Alloy | 316L SS | Ti64 | IN718 | AlSi10Mg |
|---|---|---|---|---|
| **Composition (at. %)** | $Fe_{69}$-$Cr_{18}$-$Ni_{12}$-$Mo_1$ | $Ti_{86}$-$Al_{10}$-$V_4$ | $Ni_{53}$-$Cr_{23}$-$Fe_{18}$-$Nb_3$-$Mo_2$-$Ti_1$ | $Al_{90}$-$Si_9$-$Mg_1$ |
| **Limited Phases** | LAVES_PHASE_C14 | | NBNI3, NI3TI, LAVES_PHASE_C14 | MG2SI, DIAMOND_FCC_A4 |

---

$^\dagger$ The one exception to this occurs in the Ti64-IN718 path around a normalized path distance of 0.1. This clipping of a limited phase region happens because the BPP is performed discretely at a fixed resolution so small incursions into bad or limited may occur between the discrete points, but these represent very short intervals (<~0.001 normalized path distance) and small incursions (<~0.01 mole fraction).

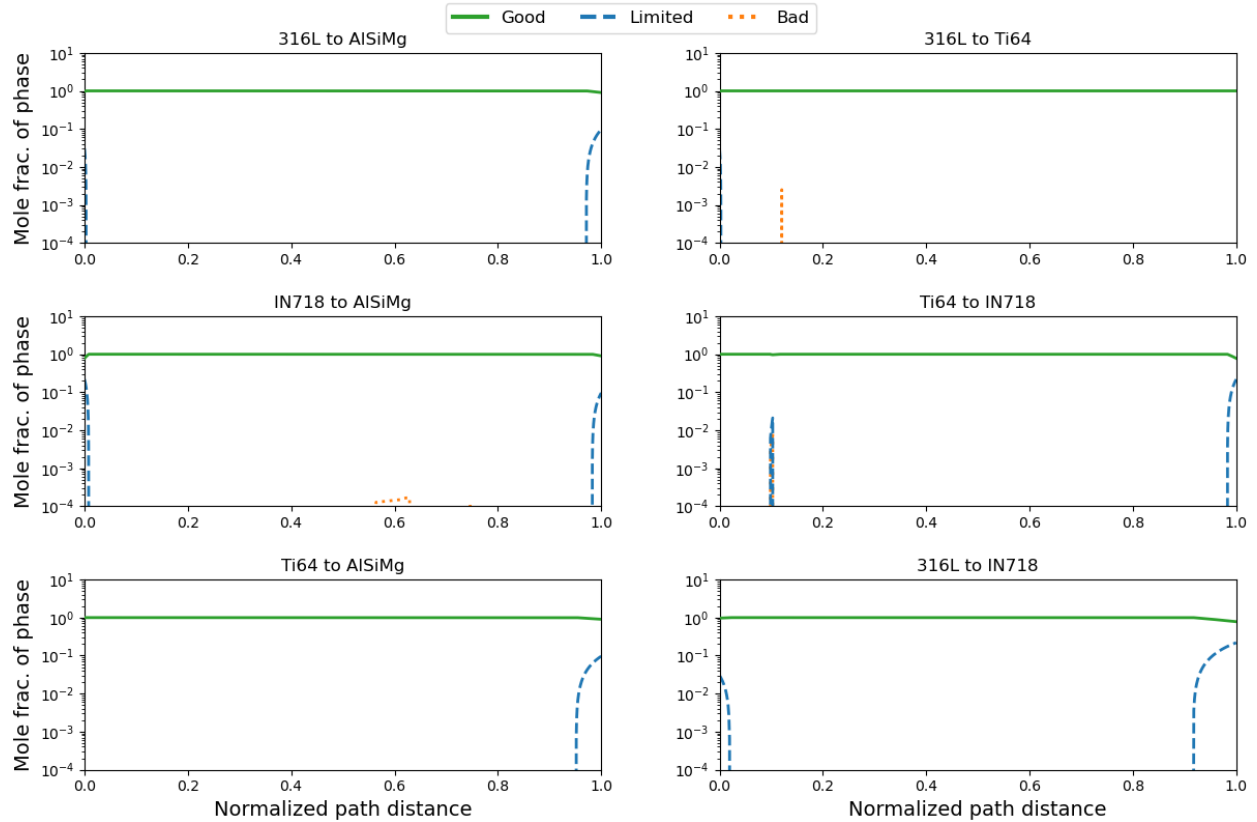

*Figure 4: The equilibrium phase fractions corresponding to the composition gradients as predicted by the TCFE9 database are shown for the six systems at 400°C. The mole fractions are summed by classification: "good" (green), "limited" (blue), and "bad" (orange). These are plotted as a function of the normalized path distance which parameterizes moving from one side of the joint to the other using the Euclidean distance traveled in composition space normalized by the total length of the path in composition space. The limited phases only appear at the ends of the gradients and in fractions less than or equal to their amount in the base alloys. The amount of bad phase is functionally zero (<$10^{-2}$) at all points.*

Figure 5 showcases the efficiency of the proposed approaches relative to the surrogate model method. The number of CALPHAD evaluations are used as a proxy for runtime to compare the performance of the two methods. This approach is taken because CALPHAD calculations take many orders of magnitude more time than other calculations in the RRT and thus dominate the total time taken. This proxy is superior when comparing measurements of the actual computation time, which are not only machine-dependent but variable with its current state. Furthermore, the time needed to perform the sampling for the surrogate model is intractable – i.e., several months would be needed to mine the data needed to generate a surrogate model in many-element dimensions – and therefore cannot be measured and used for comparison. Three separate runs (each with a different random seed) were carried out for both the strict and relaxed methods. While there is some variation in the number of GTEs between runs, the speed-up provided by the new approach ranges from $8 \times 10^2$ (Ti64 to AlSiMg) to $10^6$ (Ti64 to AlSiMg).

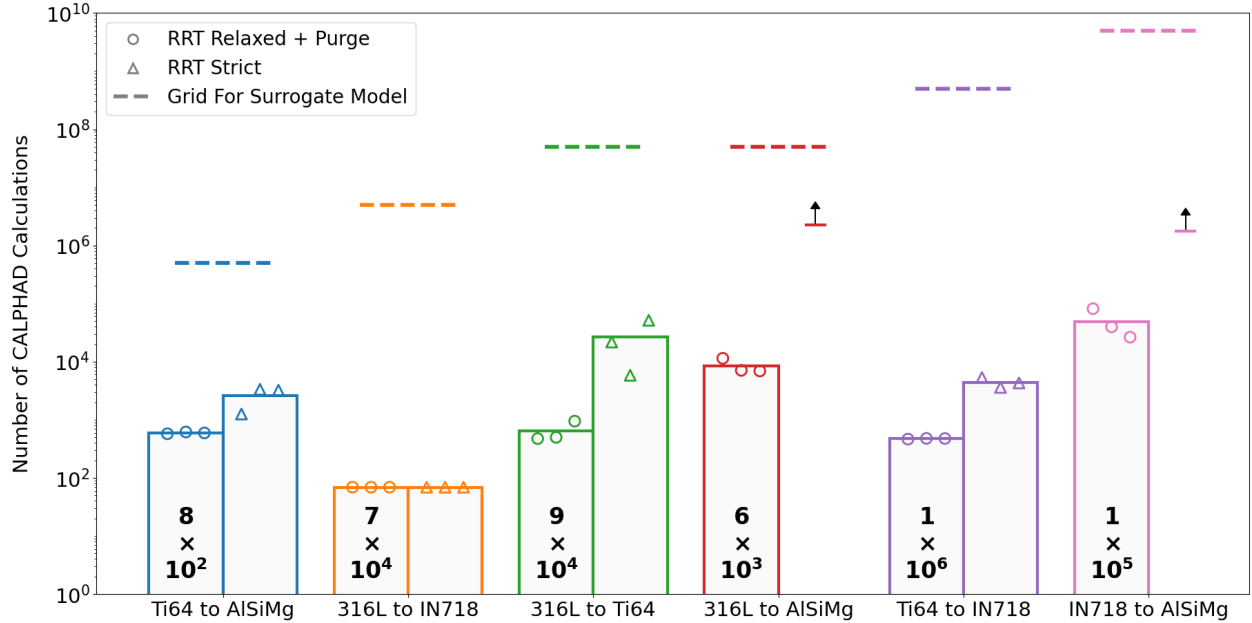

*Figure 5: Comparison of the number of CALPHAD calculations for three different path planning strategies: surrogate modeling, strict OTF, and relaxed OTF + BPP with Equation 1. The bar chart shows the mean number of CALPHAD evaluations needed to find the first feasible path using an RRT with the relaxed bad phase constraint and subsequent bad phase purge for each of the six systems. The RRT was run with three different random seeds to produce the average and the results of the individual runs are shown in the filled circles. One estimate for the number of CALPHAD calculations that would be required to train a surrogate model, $5 \times 10^d$ where d is the number of elements in the system, is shown with the dashed lines. The bolded values in each of the bars represent the speed-up over the surrogate model approach provided by this new method $\left(\frac{\text{\# of evals for surrogate model}}{\text{\# of evals for RRT+Purge}}\right)$.*

The number of CALPHAD calculations needed to find the first feasible path in Figure 5 scales favorably as the elemental dimension increases. For instance, comparing the Ti64-AlSiMg and Ti64-IN718 systems, the surrogate models differ by three orders of magnitude (5 versus 8 elements, respectively), but the strict RRT method was able to find a viable path in approximately the same number of CALPHAD queries. However, as noted above, the number of necessary GTEs is more strongly dependent on the visibility rather than the number of elements in the system. This fact is evidenced in the paths between AlSiMg and any of the other three alloys. Due to its strong affinity to form intermetallics, Al-rich regions have narrow corridors for path planning, and the strict approach was unable to find paths from AlSiMg to both 316L and IN718 in a reasonable time. However, by increasing the allowable fraction of deleterious phases, the relaxed + BPP method found viable paths for both systems with comparable GTEs to the other alloy pairs – despite planning in 7 and 9 elemental dimensions, respectively.

Three runs for each method were attempted for each alloy pair to assess variations in the number of CALPHAD calculations. However, they do not necessarily represent the true statistical distribution of possible paths. Since the RRT sampling occurs in continuous space, the probability that any two paths are identical (same path length, passing through the same phases, passing through the same compositions within a phase, etc.), or overlap for even a single point, is effectively zero. The number of paths needed to approximate the distribution will also depend on system visibility and the relaxed constraint. For this reason, it was considered beyond the scope of this work to determine how many paths should be attempted for an alloy pair.

Nevertheless, comparisons can be made between the composition profiles for the paths in Figure 5 (circle and triangle symbols). Supplementary Figure 2 summarizes the similarity results for the 316L-Ti64 pair, using Dynamic Time Warping (DTW) [22]; comparisons for the other five pairs can be found in Supplementary Figure 3. This method assesses similarity between any two paths in the same space and outputs a numerical value corresponding to the sum of squared differences between the path points after optimally matching the sequences. Thus, in DTW a lower number indicates similarity. In general, the scores within a set of paths (i.e., comparing amongst strict or relaxed paths for a given alloy pair) have comparable values, mirroring the similarities in their respective number of CALPHAD calculations. Interestingly, comparisons between the two different methods do not have a uniform trend across the alloy pairs, exhibiting lower, higher, and comparable DTW scores (see Supplementary Figure 3).

Figure 6 includes the composition profile of four 316L-Ti64 paths (two strict and two relaxed) to provide the reader context between the value of a score and the composition profiles. For the two strict paths, the composition profiles are similar and exhibit a strong peak in V in the middle. However, the two relaxed paths differ from both the strict paths and one another. V is present in much lower fractions, and Cr takes its place as the dominant element in the middle of the path. In addition, one of the relaxed paths reduces the amount of Cr and increases the amount of Mo in the middle of the path. These differences are reflected in their respective scores, but it is also important to evaluate the phases present.

Figure 7 provides additional context for the reader, showing the phase information at a single point where the composition differs the most amongst the four paths (0.6 normalized path distance). The fraction of non-good phases in the relaxed + BPP paths are both less than $10^{-9}$ due to the BPP post-processing. In contrast, both strict paths contain some amount of the Laves phase, with a similar composition across the runs, because it is a "limited" phase (allowable up to a certain fraction). Interestingly, the phase fractions of "good" phases differ quite substantially among the paths at this point. Both strict paths have two BCC phases (one V-rich and one Mo-rich), whereas the relaxed paths possess Mo- and Cr-rich BCC phases; an HCP phase is only found in one of the paths (strict 0). This is only one example but highlights the complexity of path planning in high dimensions and how different phases can emerge/be suppressed with modest changes in composition.

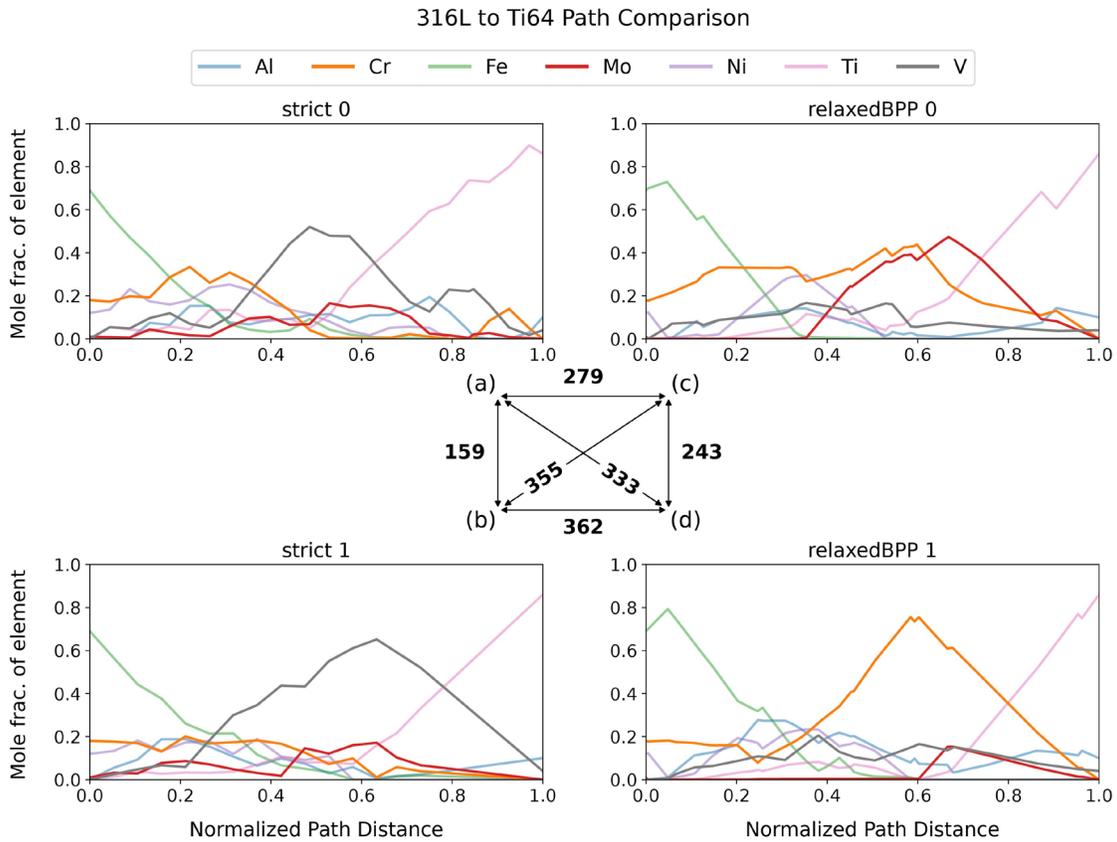

*Figure 6: Comparison of composition profiles between 316L and Ti64 with the strict or relaxed + BPP condition and either random condition 0 or 1. The DTW values corresponding to the six unique pairs between the four runs are shown in bold. While paths (a-d) do not contain deleterious phases, the strict paths (a&b) have a significant amount of "limited phases" within the interior of the path. This outcome directly results from the BPP condition, which removes limited and bad phases from points along the path that are not associated with the start/end conditions. It could be argued that some fraction of limited phases would be desirable within the gradient, which could be specified in the RRT constraints (i.e., no more than a specified fraction of limited phases) and/or cost function (i.e., target a specific fraction of limited phases).*

### Strict 0

| Phase | Phase Mole % | Al | Cr | Fe | Mo | Ni | Ti | V |
|---|---|---|---|---|---|---|---|---|
| Overall | | 9.6 | 0.2 | 1.3 | 15.2 | 2.5 | 29.4 | 41.7 |
| BCC_A2#1 | 20.2 | 2.1 | 0.0 | 0.0 | 70.6 | 0.0 | 26.3 | 1.0 |
| BCC_A2#2 | 57.4 | 8.4 | 0.3 | 2.3 | 1.5 | 0.0 | 15.8 | 71.7 |
| HCP_A3#1 | 13.8 | 20.3 | 0.0 | 0.0 | 0.4 | 0.0 | 76.5 | 2.7 |
| LAVES_PHASE_C14#1 | 8.5 | 18.8 | 0.0 | 0.0 | 0.0 | 29.1 | 52.1 | 0.0 |

### Relaxed + BPP 0

| Phase | Phase Mole % | Al | Cr | Fe | Mo | Ni | Ti | V |
|---|---|---|---|---|---|---|---|---|
| Overall | | 1.6 | 43.2 | 0.0 | 37.0 | 0.0 | 12.6 | 5.6 |
| BCC_A2#1 | 50.3 | 1.5 | 0.5 | 0.0 | 73.3 | 0.0 | 24.7 | 0.0 |
| BCC_A2#2 | 49.7 | 1.8 | 86.4 | 0.0 | 0.2 | 0.0 | 0.3 | 11.3 |
| LAVES_PHASE_C14#1 | 0.0 | 0.1 | 55.5 | 11.0 | 1.3 | 0.0 | 32.1 | 0.0 |

### Strict 1

| Phase | Phase Mole % | Al | Cr | Fe | Mo | Ni | Ti | V |
|---|---|---|---|---|---|---|---|---|
| Overall | | 3.9 | 6.5 | 0.0 | 16.4 | 0.0 | 10.5 | 62.6 |
| BCC_A2#1 | 21.2 | 0.8 | 0.0 | 0.0 | 73.2 | 0.0 | 25.1 | 0.8 |
| BCC_A2#2 | 78.7 | 4.7 | 8.3 | 0.0 | 1.1 | 0.0 | 6.5 | 79.4 |
| LAVES_PHASE_C14#1 | 0.1 | 17.8 | 0.0 | 0.0 | 0.0 | 31.0 | 51.2 | 0.0 |

### Relaxed + BPP 1

| Phase | Phase Mole % | Al | Cr | Fe | Mo | Ni | Ti | V |
|---|---|---|---|---|---|---|---|---|
| Overall | | 8.0 | 75.0 | 0.2 | 0.4 | 0.0 | 0.1 | 16.3 |
| BCC_A2#1 | 100.0 | 8.0 | 75.0 | 0.2 | 0.4 | 0.0 | 0.1 | 16.3 |
| CR3SI#1 | 0.0 | 25.0 | 1.4 | 0.0 | 73.7 | 0.0 | 0.0 | 0.0 |

*Figure 7: Equilibrium data for 316L to Ti64 gradients at 0.6 normalized path distance.*

## 3.4 Dimensional Scaling

Given this paper's focus on the utility of OTF methods, it is worth discussing their dimensional scaling against surrogate models. The shortest possible path between two alloys in composition space is a straight line, and thus the number of CALPHAD calculations needed to find a viable path should be comparable to the distance between the two composition vectors of the alloy pair divided by the step size for collision checking (0.01, in our case). This point is observed in 316L-IN718, which has an end-to-end distance of ~0.66 and path planning required 70 CALPHAD calculations. However, as seen for the other alloy pairs in this work, deleterious phases act as obstacles in composition space and make it challenging for the strict RRT to navigate. This issue is addressed by relaxing the allowable fraction of deleterious phases, and the RRT is able to find a viable path in $10^3$-$10^4$ CALPHAD calculations.

These calculation numbers are still 1-2 orders of magnitude larger than the composition distance between the two alloys but will almost always be more efficient than a surrogate model. Because the path is a continuous line, its scaling depends on both the distance and obstacles between two alloys. In contrast, the surrogate model requires sampling a hypervolume. This trend is highlighted in Figure 8, where the OTF method becomes increasingly more efficient than the surrogate model with increasing dimension. The reciprocal of the y-axis in Figure 8 can also be interpreted as how many paths can be generated via these OTF methods before it is more practical to generate a surrogate model. Clearly, for a path between two real alloys, an OTF approach is more efficient.

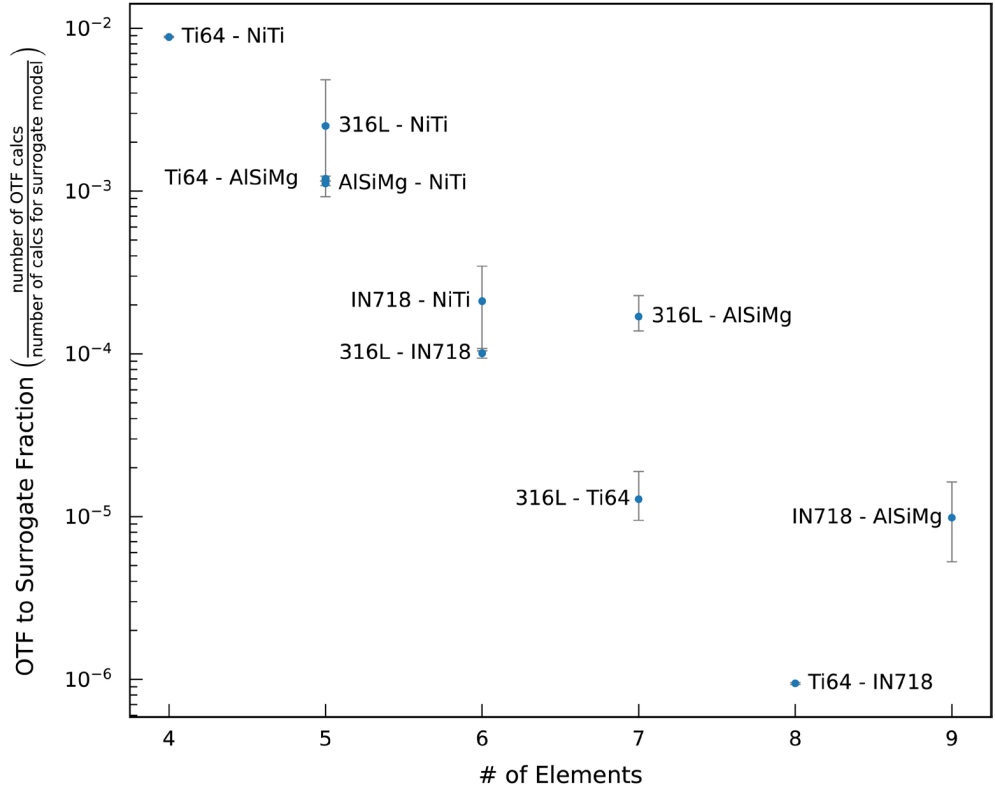

*Figure 8: Correlation between the number of elements in the system and the number of CALPHAD calculations needed to find the first feasible path (with relaxed + BPP scheme) as a fraction of the number of calculations needed to construct the complete surrogate model. The paths shown are the ten unique pairs of five alloys (the four alloys described previously along with NiTi). In this case, the 316L – IN718 path was performed the same as the others rather than using the straight-line shortcut.*

The reader may question the utility of only finding the first feasible path, citing that the composition profile in Figure 2a is more experimentally desirable than that in Figure 2c. To this end, we have extended the combined OTF/BPP approach allowing it to be implemented with RRT* and construct paths that optimize any cost function that can be computed from properties of the individual equilibrium phases (e.g., path length, value of a property[‡], gradient of a property, *but not distance to bad phases*). In this adapted strategy, the BPP is performed on-the-fly after each sample (assuming the sampled point lies in a partially bad region) rather than after path planning. The CALPHAD result for a given sample contains all the information needed to compute the purged composition and its new cost. Thus, any attempted points or edge connections that pass through the partially bad region are replaced with their transformed counterparts during the collision checking step without the need for additional CALPHAD queries. This allows the approach to maintain the improved visibility afforded by the relaxed constraints while still calculating, and thus optimizing, the true path that will meet the strict constraints (the purged path) in real time.

---

[‡] Importantly, the CALPHAD method calculates overall system properties as a function of the properties of the individual phases meaning the extension can be applied for any CALPHAD-derived property

This extended approach was used to plan optimized paths from Ti64 to AlSiMg and the results are compared to those from a surrogate model-based approach in Figure 9. The OTF approach achieves more optimal (lower cost) paths than the surrogate model approach when run for the same number of collision checks (CC's), and reasonably similar costs even when the surrogate model is used for two orders of magnitude more CC's. This improvement is due to the OTF approach performing the BPP on-the-fly allowing it to know the true cost of the path while planning, whereas the surrogate model can only optimize the path based on the cost of the unpurged path.

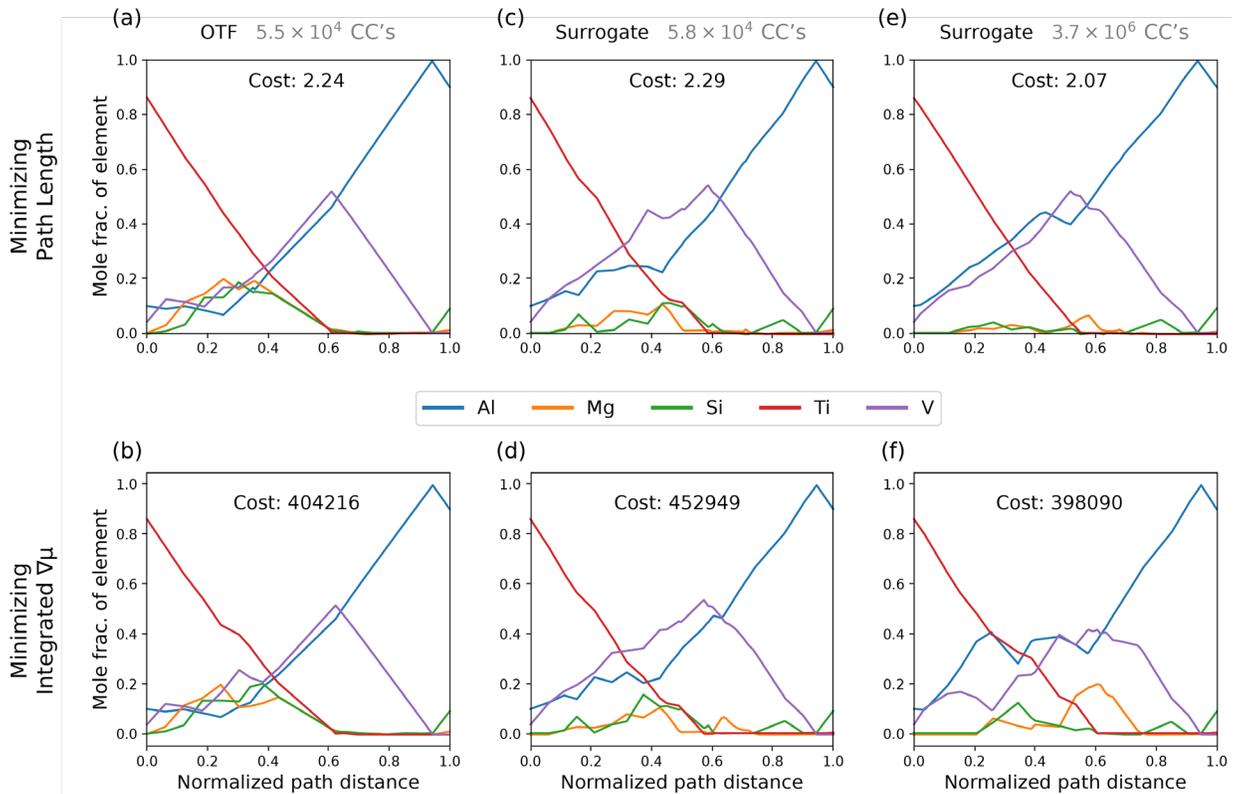

Figure 9: The composition profiles of paths planned between Ti64 and AlSiMg planned using RRT* are plotted as functions of normalized path distance. The extended OTF approach (a, b), run for a modest, $5.5 \times 10^4$ collision checks (CC's) and the surrogate model approach, run for both $5.8 \times 10^4$ (c, d) and $3.7 \times 10^6$ (e, f) CC's, were used to optimize paths with respect to two separate cost functions: path length (upper row) and the integrated chemical potential gradient (lower row). The integrated chemical potential gradient is meant to serve as an indicator for the tendency of a composition gradient to undergo diffusion, with paths having higher values being more susceptible to composition changes induced by diffusion due to a higher driving force. Unlike the extended OTF approach, the surrogate model approach could only apply the BPP method after planning. These runs differ from those shown in Figure 2a as here the BPP removes the limited phases as well.

In general, the speedup afforded by the OTF approach over the surrogate model will be less substantial for the case of optimal path planning than it is for first feasible planning, as the number of collision checks required to make a near-optimal[§] path will almost certainly be larger than the number needed to find the first feasible path (while the number of calculations to construct the surrogate model remains fixed). However, significant gains can still be expected for many-element systems. The results of Figure 9 show that the costs of the OTF produced path are comparable to that of a surrogate model produced path even as the number of collision checks used in the latter ($3.7 \times 10^6$) approaches the number of calculations used to construct the surrogate model ($4.6 \times 10^6$). The extension of the OTF approach to optimized planning is promising, but it is important to reiterate the main motivation for using the OTF approach is because for many of the systems explored in this work ($\geq 6$ elements), construction of the complete surrogate model simply cannot be performed in a reasonable time (> multiple weeks).

# 4  Conclusion

This work has focused primarily on developing a computational method capable of constructing isothermal composition paths between two compositions such that the corresponding equilibrium phases meet prescribed criteria. The major takeaways are as follows:

1. Surrogate models scale poorly when designing functional gradients that account for every element in the alloy systems of interest.
2. On-the-fly sampling can reduce the number of CALPHAD queries by orders of magnitude compared to surrogate models.
3. The efficiency of on-the-fly sampling can be further enhanced by reducing the number of collision checks that the path planning algorithm must make. Here, the number of collision checks was reduced by relaxing the constraints (i.e., allowable fraction of bad phases).
4. While this strategy appears to be at odds with gradient design goals, it can be remedied by making use of phase equilibria properties (i.e., the Lever Rule). A conjecture is provided to demonstrate this concept is generalizable, and we have yet to find an example where it does not hold.
5. We have also demonstrated that this approach can be extended to the challenge of designing gradients that optimize a particular cost function such as path length and integrated $\nabla \mu$ by using the RRT* algorithm and performing the BPP operation on-the-fly. In fact, this can be done with any cost function that can be calculated from CALPHAD-derived properties of the equilibrium phases.

---

[§] Given that RRT* is only asymptotically optimal, in many instances the path can always be made more optimal with more iterations and thus more collision checks. In this sense, the OTF approach would never "break-even" over the surrogate model because an infinite number of collision checks would be required. Hence, we use the term "near-optimal" to describe a path which is close enough to the true optimal path to be deemed sufficient by the user.

This work also identifies several opportunities for future research:

1. A hybrid sampler, which stores the results generated by on-the-fly sampling and uses them to predict the results of future queries if they lie sufficiently close to the previous known results, could provide the best of both on-the-fly sampling and traditional surrogate modeling. The algorithmic foundations of these hybrid methods for collision checking have already been developed in the path planning community [23].
2. The gradients considered in this work were generated with an assumption that the fraction of an element can be varied independently. Experimental constraints, such as the number of powder hoppers, could be integrated into this framework for optimizing the choice of powders both during and after path planning
3. Other considerations stemming from the manufacturing process, such as limitations on the spatiotemporal resolution of composition changes and variability in targeted vs realized compositions, can be accounted for in the design phase in the form of additional constraints or alterations to the cost function.
4. The concepts in this work have focused on relative phase fractions but could be extended to other gradient considerations such as designing for CTE mismatch, elastic modulus, and other mechanical properties.

# 5 Acknowledgements

This work is supported by NASA grant number ECF 80NSSC21K1810, and the Department of Defense through the NDSEG fellowship. The authors would like to thank John Reidy for his input, which provided a pivotal impetus for the application of the OTF approach to RRT algorithms.

# Appendices

## Appendix A: Bad Phase Purge Conjecture

In order to accelerate the creation of feasible paths we leverage the following conjecture:

1. Let $Z$ ($\Delta^{d-1}$, standard simplex, where $d$ is the number of elements in the system) denote the composition space
2. Let $z_{start} \in Z$ be the starting composition and $z_{end} \in Z$ be the ending composition
3. Let a composition gradient be represented by $\sigma: [0,1] \to Z$ a continuous function such that $\sigma(0) = z_{start}$ and $\sigma(1) = z_{end}$
4. Let $f_{undesirable}: Z \to [0,1]$ be a function that maps a composition to the total amount of undesirable phases present in equilibrium (at the isothermal temperature)

We conjecture:

$$f_{undesirable}(z_{start}) = 0, f_{undesirable}(z_{end}) = 0, \text{ and } f_{undesirable}(\sigma(\alpha)) < 1 \; \forall \; \alpha \in [0,1]$$

$$\Rightarrow$$

$$\exists \; \sigma^*: [0,1] \to Z, s.t. \; \sigma^* \text{ is continuous}, \sigma^*(0) = z_{start} \text{ and } \sigma^*(1) = z_{end}$$

$$\text{and } f_{undesirable}(\sigma^*(\alpha)) = 0 \; \forall \; \alpha \in [0,1]$$

additionally,

$\sigma$ can be continuously deformed into $\sigma^*$ while always satisfying $f_{undesirable}(\;\;) < 1$

In addition to establishing the existence of bad phase free paths under these conditions, we can in fact construct such a path by applying the following transformation to the original path:

$$\sigma^*(\alpha) = H_2(\sigma(\alpha)) \left[ \frac{1}{(H_1(\sigma(\alpha)) \cdot \vec{V}_{good})} (H_1(\sigma(\alpha)) \odot \vec{V}_{good}) \right]$$

Where $H_1(z): \Delta^{d-1} \to \Delta^{|S|}$ maps overall composition to phase fraction, $H_2(z): \Delta^{d-1} \to \mathcal{M}_{d \times |S|}(\mathbb{R})$ maps overall composition to a matrix of the compositions of the individual phases in equilibrium, $\vec{V}_{good} \in \mathbb{R}^{|S|}$ $s.t. \forall \; i \in [\![1, |S|]\!] \; v_i \begin{cases} = 1, \text{ if phase i is good} \\ = 0, \text{ if phase i is bad} \end{cases}$, $\cdot$ is the dot product of two vectors, $\odot$ is the Hadamard product or elementwise multiplication of two vectors, and $|S|$ is the total number of distinct phases.

This transformation is applied pointwise to a discretized version of the final path, since we can only sample the value of $H_1(z)$ and $H_2(z)$ at points by performing a CALPHAD calculation. The more finely the initial path is discretized the better the pointwise transformation will approximate the true path.

It is relatively straightforward to show that the resulting path contains no bad phases (see Supplemental). However, it remains to be rigorously proven that the transformed path is continuous (does not contain discrete jumps in composition). Despite this, we have yet to find a counterexample to the conjecture.

**Supplementary Figures**

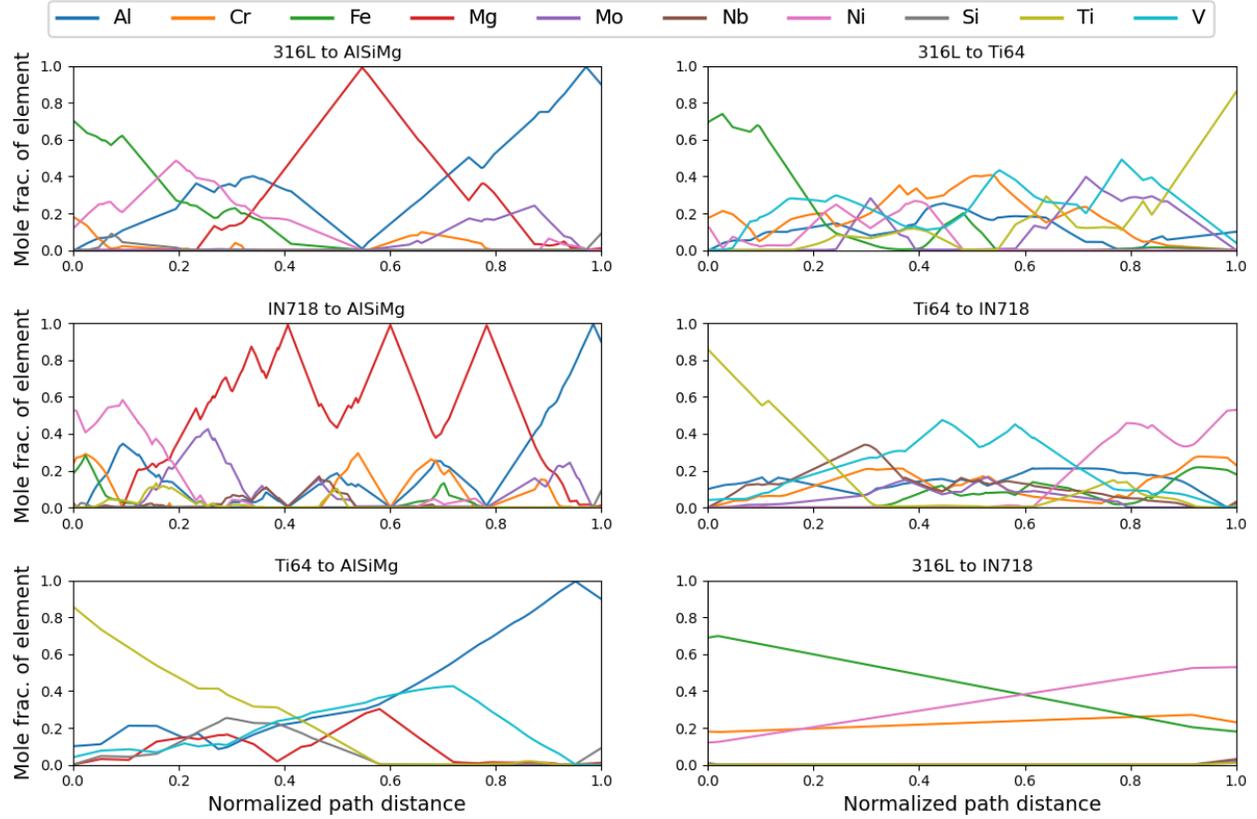

*Supplementary Figure 1: The composition gradients resulting from the relaxed + BPP approach are shown for the six systems at 400°C. These are plotted as a function of the normalized path distance which parameterizes moving from one side of the joint to the other using the Euclidean distance traveled in composition space normalized by the total length of the path in composition space.*

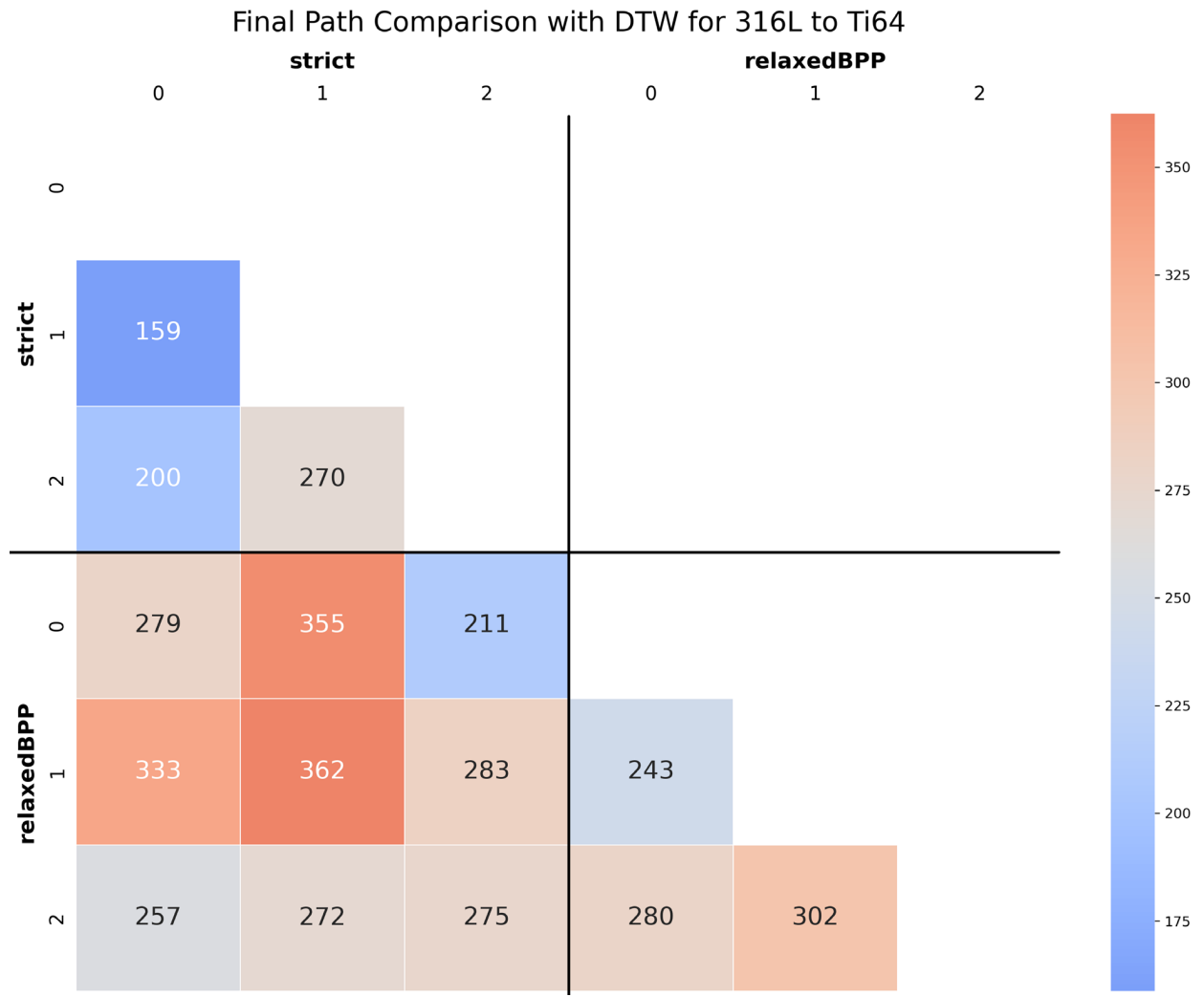

Supplementary Figure 2: Heatmap showing the individual DTW values for 316L to Ti64 for the two conditions (strict and relaxed + BPP) and the three random seeds.

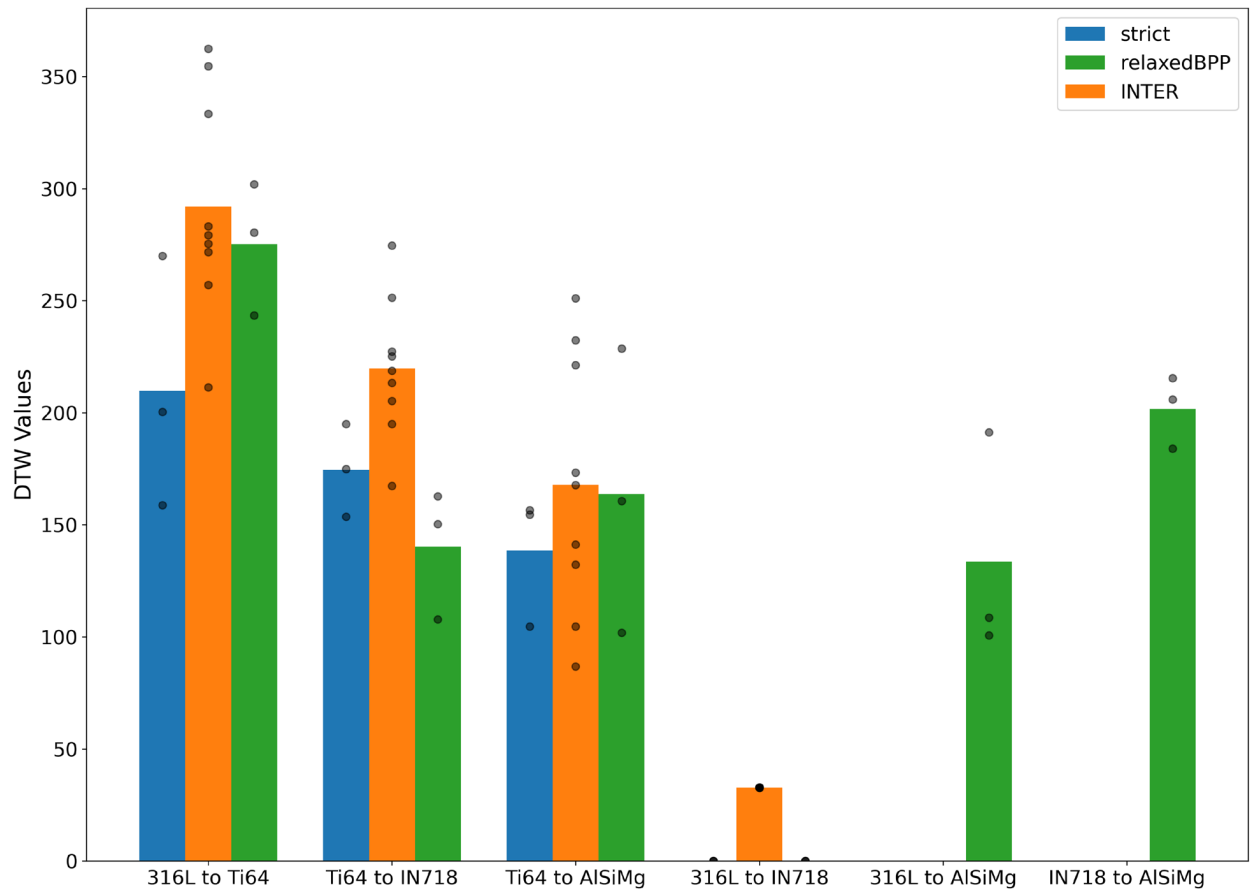

*Supplementary Figure 3: The bar chart shows the average of the dynamic time warping (DTW) values for each of the six paths grouped by comparisons within the same condition (strict and relaxed + BPP) and those between conditions (INTER). The individual values are shown with the black circles.*

## Supplementary Notes

### Partial Proof of Bad Phase Purge Conjecture:

Two things to show:

1. All transformed points are entirely not bad (i.e., completely good): $H_1(\sigma^*(\alpha)) \cdot \vec{V}_{bad} = 0$
2. The transformed path, $\sigma^*(\alpha)$, is continuous:

We can prove (1):

We have that

$$H_1(\sigma^*(\alpha)) = \frac{1}{(H_1(\sigma(\alpha)) \cdot \vec{V}_{good})} (H_1(\sigma(\alpha)) \odot \vec{V}_{good})$$

Thus,

$$H_1(\sigma^*(\alpha)) \cdot \vec{V}_{bad} = \frac{1}{(H_1(\sigma(\alpha)) \cdot \vec{V}_{good})} (H_1(\sigma(\alpha)) \odot \vec{V}_{good}) \cdot \vec{V}_{bad}$$

By the properties of the Hadamard product and dot product the following relation is true[**]: $(\vec{a} \odot \vec{b}) \cdot \vec{c} = \vec{a} \cdot (\vec{b} \odot \vec{c})$, and using this we find:

$$\frac{1}{(H_1(\sigma(\alpha)) \cdot \vec{V}_{good})} (H_1(\sigma(\alpha)) \odot \vec{V}_{good}) \cdot \vec{V}_{bad} = \frac{1}{(H_1(\sigma(\alpha)) \cdot \vec{V}_{good})} (H_1(\sigma(\alpha)) \cdot (\vec{V}_{good} \odot \vec{V}_{bad}))$$

By their definition we have $\vec{V}_{good} \odot \vec{V}_{bad} = \vec{0}$

$$H_1(\sigma^*(\alpha)) \cdot \vec{V}_{bad} = \frac{1}{(H_1(\sigma(\alpha)) \cdot \vec{V}_{good})} (H_1(\sigma(\alpha)) \cdot \vec{0}) = \frac{1}{(H_1(\sigma(\alpha)) \cdot \vec{V}_{good})} * 0 = 0$$

Similarly, we can show

$$H_1(\sigma^*(\alpha)) \cdot \vec{V}_{good} =$$

---

[**] $(\vec{a} \odot \vec{b}) \cdot \vec{c} = D_c \vec{a} \cdot \vec{b} = (D_c \vec{a})^T \vec{b} = \vec{a}^T D_c^T \vec{b} = \vec{a}^T D_c \vec{b} = \vec{a}^T \cdot (D_c \vec{b}) = \vec{a} \cdot (\vec{c} \odot \vec{b}) = \vec{a} \cdot (\vec{b} \odot \vec{c})$ where $D_c$ is a diagonal matrix with the vector $\vec{c}$ as its main diagonal

$$\frac{1}{\left(H_1(\sigma(\alpha))\cdot \vec{V}_{good}\right)}\left(H_1(\sigma(\alpha))\odot \vec{V}_{good}\right)\cdot \vec{V}_{good}$$

$$=\frac{1}{\left(H_1(\sigma(\alpha))\cdot \vec{V}_{good}\right)}\left(H_1(\sigma(\alpha))\cdot \left(\vec{V}_{good}\odot \vec{V}_{good}\right)\right)$$

By their definition we have $\vec{V}_{good}\odot \vec{V}_{good} = \vec{V}_{good}$

$$=\frac{1}{\left(H_1(\sigma(\alpha))\cdot \vec{V}_{good}\right)}\left(H_1(\sigma(\alpha))\cdot \vec{V}_{good}\right) = 1$$

Additionally, since $z_{start}$ and $z_{end}$ are both 100% good (as stated in the conditions of the conjecture) they remain unchanged by the bad phase purge operation and so $\sigma^*(0) = z_{start}$ and $\sigma^*(1) = z_{end}$.

We have been unable to prove (2). Thus, it remains to be rigorously shown that the transformed path is continuous (does not contain discrete jumps in composition). Despite this, we have yet to find a counterexample to the conjecture.